\documentclass[runningheads]{llncs}
\usepackage[T1]{fontenc}
\usepackage{graphicx}
\usepackage{epsfig}
\usepackage{calc}
\usepackage{multicol}
\usepackage{pslatex}
\usepackage{hyperref}
\usepackage{enumitem}
\usepackage[bottom]{footmisc}
\usepackage{xcolor}
\usepackage{todonotes}
\usepackage{comment}
\usepackage{etoolbox}

\begin{document}

\title{Toward Methodical Discovery and Handling of \\ Hidden Assumptions in Complex Systems and Models}
\titlerunning{Discovering Hidden Assumptions in Systems and Models}
%

\author{David Harel\inst{1}\orcidID{0000-0001-7240-3931} \and
Uwe Aßmann\inst{2}\orcidID{0000-0002-3513-6448}  \and \\
Fabiana Fournier\inst{3}\orcidID{0000-0001-6569-1023} \and
Lior Limonad\inst{3}\orcidID{0000-0002-4784-2147} \and \\ 
Assaf Marron\inst{1}\orcidID{0000-0001-5904-5105} \and
 Smadar Szekely\inst{1}\orcidID{0000-0003-1361-1575}
}
 \institute{Department of Computer Science and Applied Mathematics, \\ Weizmann Institute of Science, Rehovot, Israel \\ \email{david.harel,assaf.marron,smadar.szekely\}@weizmann.ac.il} \\ \and 
 Technische Universität Dresden, Dresden, Germany  \email{ uwe.assmann@tu-dresden.de}  \\ \and 
 IBM Research, Haifa, Israel\email{\{fabiana,liorli\}@il.ibm.com} 
}

\authorrunning{Harel, Assmann, Fournier, Limonad, Marron, and Szekely}

\maketitle              
\begin{abstract}
Methodologies for development of complex systems and models include external reviews by domain and technology experts. Among others, such reviews can uncover undocumented built-in assumptions that may be critical for correct and safe operation or constrain applicability. 
Since such assumptions may still escape human-centered processes like reviews, agile development, and risk analyses, here, we contribute toward making this process more methodical and automatable.  
We first present a blueprint for a taxonomy and formalization of the problem.
We then show that a variety of digital artifacts of the system or model can be automatically  checked against extensive reference knowledge.
Since mimicking the breadth and depth of knowledge and skills of experts may appear unattainable, we illustrate the  basic feasibility of automation with rudimentary experiments using OpenAI's ChatGPT. 
We believe that systematic handling of this aspect of system engineering can contribute significantly to the quality and safety of complex systems and models, and to the efficiency of development projects.
We dedicate this work to Werner Damm, whose contributions to modeling and model-based  development, in industry and academia, with a special focus on safety, helped establish a solid foundation to our discipline and to the work of many scientists and professionals, including, naturally, the approaches and techniques described here.


\end{abstract}

\bibliographystyle{splncs04}

\section{Introduction}\label{sec:intro}

Many famous failures of complex systems may be related directly or indirectly to unstated assumptions.
For example, the famous Year-2000 (a.k.a Y2K) ``bug" is related to assumptions like  ``All calendar years mentioned in this system begin with `19' (or can be  represented with two digits).'' 
In 1999, the Mars Climate Orbiter spacecraft crashed because of the assumption that certain interacting software components used the same measurement units when in fact they were not.
Some published reports and analyses of the tragic Tesla car crash in 2016 relate to human user assumptions about the automated driving capabilities of the vehicle,
and/or to technical assumptions about which sensor input to rely on in which context~\cite{banks2018Tesla}. 
In 1991, the Patriot missile defense system missed an incoming missile with fatal consequences, due to  cumulative time drift resulting from data truncation.
This bug may be retrospectively associated with an assumption that the system will be reset frequently,
and/or that the speed of incoming missiles is below a certain  threshold~\cite{marshallSCIENCE1992PatriotTimingBug}. 
Environmental and biological models, e.g., with regard to global warming or Covid-19,  may also contain explicit and implicit assumptions that affect their predictions. ~\cite{medhaug2017globalWarmingHiatus,biggerstaff2022covid19Modeling,kotwal2020CovidModels}. 
In the words of ~\cite[p.11-1, col.2]{pelletier2003aboutIncorrectModels} \textit{``it is possible that incorrect models were `tuned’ to observed behavior.''}. The problem is of course general: despite thorough development and testing, many operational systems experience failures and near failures that are rooted in hidden assumptions. 

Present system engineering methodologies and tools~\cite{buede2016Systemengineering,bialy2017softwareEngineeringMBD,masso2020riskMgtReview,kramerDamm2020identificationHazards} 
of course attempt to address some of these issues in elements of:
\begin{itemize}
    \item requirements engineering and requirement traceability, 
    \item risk analysis,
    \item testing and validation, including unit,  feature and integration testing,
    \item test-driven development methods, 
    \item agile development with repeated involvement of users and customers,
    \item Reviews of documents and code by stakeholders\footnote{We use the term stakeholders broadly, and include end users, customers/clients, developers/engineers, architects, product managers, project managers, etc.} with diverse backgrounds, 
    \end{itemize}
etc. In this paper, we are less interested in hidden assumptions that were missed because developers did not follow such practices or applied  other risky development behavior 
(as in~\cite[page 7]{charette2005softwareFailures}), 
and focus on those that despite systematic use of existing methods remained in the ``blind spot'' of developers and other stakeholders. We also note that there may be considerations that stakeholders did consider explicitly, but incorrectly dismissed them as inapplicable without documentation.

Hidden assumptions are not always a negative factor. They often contribute to simplicity, efficiency, or cost reduction in design and development. 
For example, a critical business application that under certain conditions issues an error message and stops processing, may depend on a common undocumented assumption that it is used within a business process in which a human will handle these messages.  For another example, consider useful low-cost electronic devices that are obviously not designed to operate in harsh conditions or in bright sunlight, but for which this fact is not documented. 
Thus, since deliberating an elicited hidden assumption may itself induce cost and complications, one should consider the benefit of bringing it to the forefront. 
For example, recognizing that a business application is dependent on a human operator may lead to actions like documenting the process and training the operators, or enhancing the system to handle such errors automatically; alerting users to limitations of low cost gadgets can promote the acquisition of alternative solutions; etc.  

More generally, handling discovered hidden assumption includes documenting the missing details and alerting affected parties, or modifying the system and/or its testing such that the assumption is no longer valid or is replaced by other documented assumptions. Furthermore, each discovery of a hidden assumption should cause developers to evaluate whether this discovery should have been occurred earlier, leading to potential improvements in the development process. 

While in the case of complex safety-critical systems the importance of the issues discussed here is manifest, the proposed methods and tools can apply in diverse contexts, including commercial and scientific, enterprise and personal, operational control and information processing, etc.

In Section~\ref{sec:types}  we contribute several perspectives and categories for studying hidden assumptions, toward constructing a systematic taxonomy. Then, in Section~\ref{sec:discovery}, we use such categories to outline methodical and automatic processes for discovery of hidden assumptions. In Section~\ref{sec:ChatGPT}, we support and illustrate the argument that automation of this process, which when done by humans requires extensive knowledge and advanced analytical skills, is at all possible. We do this via  rudimentary experiments using OpenAI's ChatGPT to compare documents, analyze code, and infer ontologies in the context of general and domain-specific knowledge.

The approaches discussed here tie directly into ideas and concepts introduced and/or developed by Werner Damm and colleagues including scenario-based programming (SBP)~\cite{dammHarel2001LSC} and contract-based development of cyber-physical systems~\cite{Dammsangiovanni2012contractCPS,DammBenvenisteHenzingerEtAl2018contracts} and a variety of related topics in developing and verifying complex systems\cite{damm2018formalTrafficSequenceCharts,kramerDamm2020identificationHazards,bienmullerDamm2000statemateVerification}. 
A common theme of this view of software and system engineering is the combination of separation of concerns and comprehensive coverage. Every requirement, expectation, constraint, etc., whether executable or not, deserves a concise specification, enabling full awareness of stakeholders and prioritization of systematic handling in implementing it or in using components that implement it. 

In the spirit of the wise computing vision~\cite{harel2018wiseComp}, researching and automating this aspect of requirements analysis and quality assurance can enhance the quality and safety of complex systems and the efficiency of development projects.

\section{Perspectives on the types of Hidden Assumptions}\label{sec:types}

\begin{figure*}[ht]
    \centering
    \includegraphics[width=\linewidth]{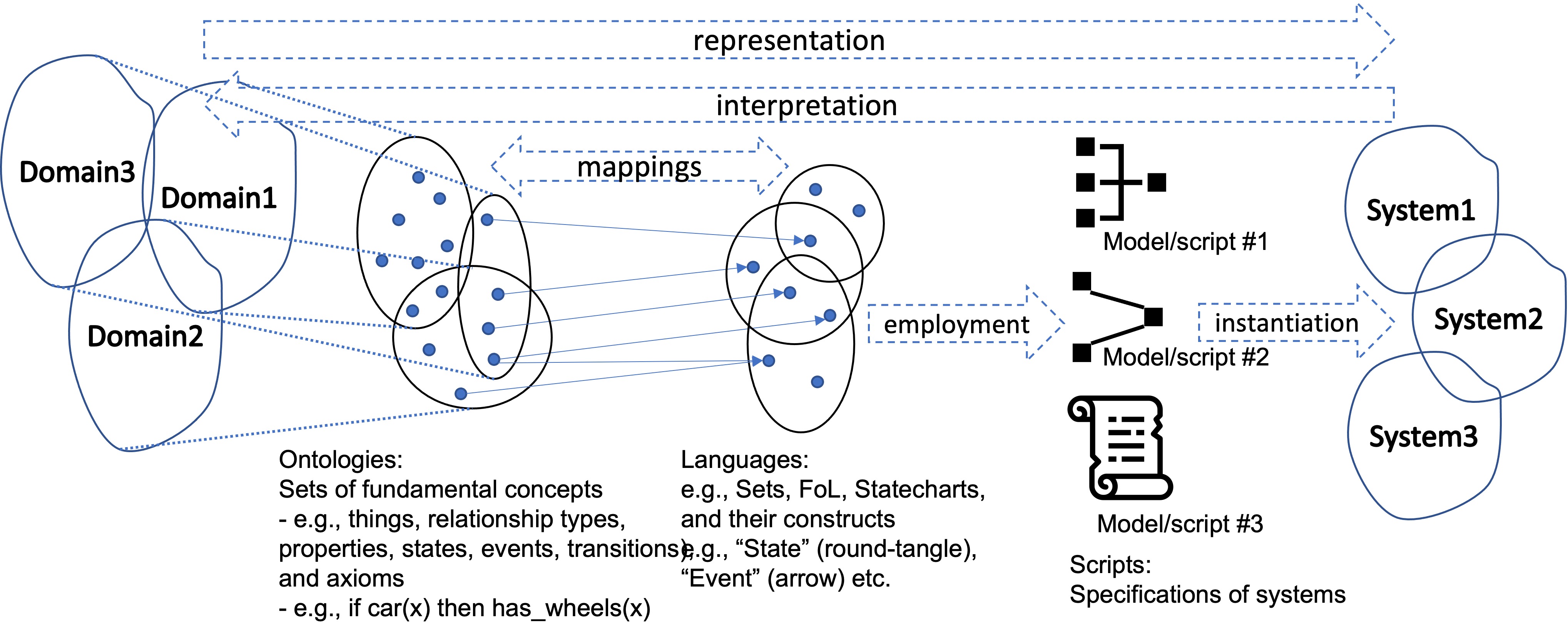}
    \caption{\textbf{From real-world problem domains to systems:} Left to right (representation): application domains, formalization of concepts within these domains, representation of the formalization as entities in specification languages, programs/scripts/specifications that encode systems structure and behavior, systems. The inverse direction reflects the interpretation of systems and their specifications as real-world effects. Such artifacts can be scanned for indications of hidden assumptions at various levels of abstraction.}
    \label{fig:domains-to-systems}
\end{figure*}

As illustrated by Lewis Carroll (Charles Lutwidge Dodgson) in ``The Hunting of the Snark'', when looking for something, it may be desirable to have a description of the object of the search, elusive as it may be.   
Most generally, we consider a hidden assumption to be a description of a system property or behavior that by its nature would fit in a requirements document or a similar object, and affects system development, testing, or use, but is not followed up on completely, because affected stakeholders are not aware of it.
Toward a more specific definition, below we offer several perspectives, 
or views, for describing and categorizing such hidden assumptions.
Figure~\ref{fig:domains-to-systems} provides a high-level ontological context, highlighting the fact that a working system, concrete and real as it may be, depends on multiple stages of abstracted specifications of the relevant real-world entities.
Furthermore, these specifications are digital and examining them for signs of hidden assumptions can be automated. 
For more details about domain, model and metamodel ontologies see~\cite{assmann-ontologies-and-mda,atkinson2003mddOntology}.

\subsection{A domain-dependent perspective}\label{sec:domainKnowledgePerspective}

Many hidden assumptions are based on the nature of the application. 
For example, 
in describing road intersections for an autonomous vehicle (AV), excluding a configuration that exists somewhere in the world
may reflect a hidden assumption that it does not exist in the planned use cases. 
Domain-specific assumptions may  be induced by the chosen abstraction levels,  implying that  different entities within a  category should be treated uniformly, regardless of their differences. For example, specifying for an AV the concept of road lanes without reference to lane widths may imply the hidden assumption that the same behaviors hold for all lane widths. Such abstractions  can  range from highest-level use cases of the system, through function modules, to fine-grain physics of the nuts and bolts, literally, as well as electronics, in the system and its environment.  

Domain-dependent assumptions often constitute ontological axioms. For example,  typical rules of behavior of an AV in a city, expect that all road users move only in two dimensions, and do not fly. Such ontological axioms may also emerge from the language used. E.g., interior design for a home may depict a ceiling fan in the middle of a room; the existence of a ceiling is implied.   

\subsection{A stakeholder-centric perspective of the concept of being hidden} 

We offer the following broad classification of discovered hidden assumptions, based on who might not be aware of them: 
\begin{itemize}
\item Not documented, but well known: The requirement, constraint, etc., is not documented, but, if asked, stakeholders of all roles would describe correctly how the system relates to the issue at hand; 
\item Assumed, and specified explicitly, but unknown to some stakeholders: The requirement is specified in code or documents that are not accessible to certain stakeholders (like customers) who should be aware of it. 
\item  Undocumented and totally hidden from many relevant stakeholders: despite their diligent joint work, developers, system engineers, customers, and users have not specified and are not aware of a certain assumption that was built into the system.
\end{itemize}

Awareness of such categories can suggest discovery techniques (See Section~\ref{sec:discovery}), like focusing on esoteric area in the problem domain or in the development process. 

\subsection{Specification-structure perspective}\label{sec:specPerspective}

In this perspective, we examine hidden assumptions that emanate from inherent traits and effects of the overall structure of the used formalism. Such formalisms include classical state machines,  statecharts, Petri nets, class diagrams, sequence diagrams and live sequence charts (LSC), temporal logic,  models in languages like Z notation and VDM, models of databases, MATLAB Simulink models, activity and use case diagrams, component diagrams,  high-level architecture diagrams, etc. Section~\ref{sec:langPerspective} discusses a programming language perspective. 
 
Such formal specifications often represent intentions compactly using named entities. For example, in modeling an AV, one would define system states and conditions, like ``engine on'' vs ``engine off'' or the color of a traffic light;  actions, like ``stop'', ``start'', or ``slow down''; events, like ``Police person signaled ego AV to pull over'', etc. Below are examples of features of formal models that may induce hidden assumptions, especially when combined with such abstract representations.  

\begin{enumerate}
\item{\textit{Abstraction, encapsulation and equivalence:} 
As discussed extensively in the context of contract-based design methodology~\cite{DammBenvenisteHenzingerEtAl2018contracts}, in many specifications, the information provided about an entity often includes its interface and excludes important conditions and shared variables it relies on, effects it generates, parallel execution dependencies, division of responsibilities, etc. Similarly,  in object-oriented design, all object instances belonging to the same class are assumed to be equivalent with regard to the class's attributes and methods; one may not readily know whether differences between subclasses of instances, and between individual instances, were or were not accounted for. If the relevant information is not proactively sought out and stated explicitly, it remains as hidden assumptions.  
} 

\item {\textit{Orthogonality and Set Disjointness:} Consider a program with parallel components or a statechart that divides a certain system state into several orthogonal regions. This may imply the assumption that all reachable combinations of substates are safe and are properly handled, and/or that unsafe combinations mysteriously become unreachable. Similarly, a modeled set may be divided into several disjoint subsets to communicate a certain abstraction, which in turn may hide important information. For example, a traffic simulation that initially distinguishes road users and stationary obstacles, clearly has to deal with objects that belong to both sets.} 

\item {\textit{Atomicity of transition:}
Specifying a process as a state machine suggests that specified transitions are atomic, i.e., uninterruptible. Since in reality, some transitions do go through interim states that may require developers' attention, the specified atomicity of the transitions carries assumptions about this handling, like that it is not required, or is handled elsewhere.}

\item{\textit{Implication by containment:} 
Formal models have clear relations of set containment. 
For example, in a statechart, an event exiting a containing state applies to all contained substates, as in the case of a transition leading to a safe state upon some critical event. This ``For all'' quantifier implies the assumption that all nested states can be interrupted and exited upon the occurrence of this event.}

\item{ \textit{Completeness:} 
Examining  a state machine or an architecture/design diagram, a domain expert might be drawn into checking the correctness of existing specifications, and not notice that certain higher level aspects are altogether missing. Unless certain entities are mentioned but are not fully specified, there is little, if anything, in a system artifact to remind an expert reviewer to check against the their personal knowledge whether certain components, behaviors, external events, or inter-entity relations  are entirely absent. 
While checking if a given complex system or a model  is complete or correct with reference to some accepted scope and abstraction level, is a question in its own right~\cite{harel2005turingBiology}, 
any gaps between a system's code and documents and an established domain knowledge, reflect hidden assumptions about the scope of the system.} 

\end{enumerate}
\subsection{A language perspective}\label{sec:langPerspective}

Hidden assumptions can arise from characteristics of the language and specification method used like syntax, semantics, and visual layout, including aspects like order and containment relations~\cite{moody-2007}. Furthermore, user interpretations of specifications are often influenced by the users' mental models and expectations~\cite{Norman2016DesignOfEveryDayThings}. 
Below we offer attributes of language dependent hidden assumptions, also viewed as \emph{specification deficits}. This list is based on an ontological paradigm adapted from~\cite{Wand1995deep-structure-WW-1995,Wand1993ont-expressiveness-WW-1993}, which  was created originally for assessing modeling grammars. 

\begin{enumerate}

\item \textit{Inexpressible:} 
A requirement that cannot be specified explicitly and directly due to language constraints. For example, in a rule-based language that is limited to statements like ``Always, when condition C holds, do action A'', one cannot readily state ``Always, after event E occurs, do Action A ''. 

\item \textit{Unclear:} A part of a specification that may be misunderstood. For example, when an inexpressible requirement is implemented indirectly, or when 
using similar names for different entities.  

\item \textit{Unintended:} A specified requirement that, due to language semantics, carries an unintended meaning. 
For example: Consider the rule ``Always, when condition C holds, do action A''. 
If action A also makes condition C false, there may be a hidden assumption that A occurs exactly once whenever the system recognizes that C is true. But, depending on timing aspects of the system, the action  may also occur several times.  

\item \textit{Opaquely encapsulated:} The discussion in Section~\ref{sec:specPerspective} about abstraction and encapsulation of application components applies also to built-in language functions.

\item \textit{Unspecified:} Information that is missing altogether. For example, the specification ``Always, after event E, do action A'' and ``Always, after event E, do action B'', does not detail whether A should be executed before, after, or concurrently with B. 

\item    \textit{Implicit:} The requirement is not stated at all, but is implied by the language or the execution foundation. For example, continuing the previous one, the language semantics may dictate that rules are  evaluated and executed to completion, one at a time, in the order in which they appear in the source files.
This is also the case with parallelism and order of execution in synchronous languages~\cite{smyth2019synchronousLanguages}, and conversely, nondeterminism in these issues as implied in other languages; still, one cannot be sure that the relevant stakeholders are fully aware of those definitions when writing or interpreting a model or a program, as part of their own actions. 

\item \textit{Perceived:} An explicit and clear specification that a human reader may misinterpret. E.g., the python statement \texttt{A = B = C}, assigning the value of C to B and then to A,  may be misinterpreted  as \texttt{A = B == C}, making A \texttt{True} if B is equal to C.

\item    \textit{Fragmented:} A requirement is stated only as a collection of sub-requirements. For example, an AV is designed per the undocumented requirement that it works only in certain geographies, and is programmed throughout only with the implied traffic laws, like driving on the right. Another example is when a requirement is implemented in two independent rules where one constitutes an overriding priority or exception to the other.  
\end{enumerate}

\subsection{Program code perspective}
In addition to the specification structure and language issues discussed in Sections~\ref{sec:specPerspective} and~\ref{sec:langPerspective}, almost every line of program code and every entity in a formal model (which is a form of program code) carries with it hidden assumptions. For example: 
\begin{enumerate}
    \item{\textit{Silent conditions, conjunctions and disjunctions:} Assume that the system checks condition C in certain contexts. For other contexts, a reviewer cannot readily tell whether the developer expects C to be always true, always  false, or irrelevant, or if perhaps, not checking C is an omission.}
    
    \item{\textit{Fixed parameters and initial values:} All constants in a system represent requirements and assumptions. However, one should distinguish between the more conspicuous assumption, about the choice of value, and the more implicit one, that this value is fixed and applies equally in all cases where it is used.}
\end{enumerate}

\subsection{Additional perspectives}

Additional categorization perspectives for hidden assumptions may be guided by   
the diverse development activities and related interests of stakeholders like those described in~\cite{Kruchten954Plus1Views}:  system logic and function view, software development process view, integration process view, physical component view, and scenarios view. 

Time is of course another relevant dimension; requirements that were so obvious that they were left undocumented in early project stages can become obscure as time passes, people change roles, needs change, etc. 

\section{An outline of methodical discovery of hidden assumptions}\label{sec:discovery}

The categorizations in Section~\ref{sec:types} suggests initial directions for creating methodologies and tools for discovery of hidden assumptions, as follows. 

\subsection{Semantic analysis and comparisons}

We wish to mimic the ability of external experts to use their knowledge, experience, and  analytical skills to examine programs, documents and demonstrations of a system and identify important omissions and hidden assumptions.  
Hence, we propose to create a mapping between generally available domain and technology knowledge on the one hand and the target system on the other. Gaps in either direction and any kind of difference in contents, perspective or level of detail will be very instructive regarding assumptions, omissions, and areas that require further attention. 

For the system being studied one would collect:  requirements documents;  architectures;  designs; models;  software code; 
test cases; environment simulation software code; user guides and operation manuals (including ``fine print'' with disclaimers and warnings);  technical specifications; sales brochures;  questions-and-answers repositories;  execution logs and even videos of the system execution in real-world tests;  trouble and error reports by QA engineers and early users, including their resolutions or reasons for closure; development meeting summaries; long term development plans and deferred feature requests; etc.   

Sources for general domain and technology knowledge would include text books, tutorials, and articles about the domain; examples of small systems, with problems and solutions;  system artifacts, as listed in the previous paragraph, for  related systems like prior versions of the target system, past solutions, competitive solutions, 
documentation of technical comparisons of such systems; regulations and industry standards; accounts of famous bugs and failures; ontology databases/schemas~\cite{zipfl2023OntologiesAV}; etc. 

This specific mapping can be readily extended beyond the available materials. For example, if a tool examining a quantitative system aspect, say, a distance between two objects, raises a valid question about the fidelity of the associated measurement or precision of related program variables, human engineers, as well as tools, may be alerted to the need to review such aspects in other parts of the application.  

\subsection{Structure-driven question generation\label{sec:StrucureDrivenDiscovery}}

One approach for structure-driven discovery of hidden assumptions and unhandled contingencies is as follows. 
For each type of broad assumptions  in the structure-driven perspective, select a set of relevant system artifacts, and state a question challenging the assumption. The questions may then be answered by a human, and in some cases, by tools based on domain and technology knowledge. For example: 

\begin{enumerate}

\item{For abstraction, encapsulation and equivalence assumptions: for any encapsulated function, check its interface definition and then ask about issues that are not directly manifested there. Candidate topics for software include performance, exception handling, resource consumption, and side effects. For hardware the missing aspects may emerge from  part features and specifications in the domain. In an object-oriented design, for each class, find pairs of subclasses or object instances, ask about their differences, and then ask whether these difference should imply refinement of superclass methods, user interaction instructions, etc.} 

\item{For Orthogonality and set disjointness assumptions: In a specification like that of a statecharts, for each state that is divided into parallel regions, run model checking to check if all composite states are reachable. If not, this can be instructive about the problem domain and application or about the design and the specification. In particular, if the unreachable composite state is an undesired one, the requirement that it is  unreachable may reflect a hidden assumption, or its coding may be highly fragmented. For set definitions, select a member of some set and and ask whether it could also belong to another set. Continuing the example of road users vs. stationary obstacles, once asked one realizes that what appears visually as a house may be a road user, when it is carried on a truck. Similarly, in a state machine, one assumes that all states are disjoint.
An examination of pairs of named states with domain expertise in mind, may elicit situations where the system may satisfy the assumed conditions for two distinct states at the same time.}

\item{For state containment: for each sub-state and each event exiting the containing state, check if exiting at that substate requires special handling that was not specified.}

\item{For atomicity of transition: in a state-based formalism, for each transition, generate questions regarding its atomicity; domain knowledge may elicit interim states that need to be specified. Also, if there is an atomic transition between two system states that differ by more than one state variable, consider whether one needs to handle the case where only some of the variables have changed. } 

\item{For Completeness: in a state-based formalism, generate questions that check whether the listed states and orthogonal regions indeed offer complete coverage. For each state, check if the incoming and outgoing transitions associated with this state are indeed the only ways to enter and exit the real-world condition represented in this formal state. For formalisms that include entities and relationships, systematically check pairs of entities, and ask whether they have relationships that have not been specified.}
\end{enumerate}

Such structure-based question generation may be intractable for large systems, due to large number of all pairs, or all possible sets, of such entities. Tools will have to employ additional techniques such as random sampling, heuristic choices of suspicious or promising subjects, automated learning from accumulated experience in such automated reviews, and incorporating interactive human assistance in the review process.
  
\subsection{Explanation-driven search} 
Much insight can be gained from human explanations and commentaries that complement established system artifacts. Automated tools can look at questions and challenges like: trouble tickets that were closed without handling because ``everything worked as expected'';  challenges by stakeholders and external reviewers that were resolved by arduous explanations of what the system can and cannot do, or where and how it should and should not be used; analysis reports following major failures; etc.  These explanations represent knowledge about the system that is not readily available to key stakeholders, and that perhaps should be better communicated and then re-evaluated.
One should note though that such human-provided explanations are not necessarily correct or complete, and involve an intermediate step of human interpretation~\cite{Norman2016DesignOfEveryDayThings}. 

\section{Experimenting with semantic analysis using ChatGPT}\label{sec:ChatGPT}

Reliance on external knowledge for automated checking of complex systems may still be far away. It may also seem that  
automating key capabilities of knowledgeable, smart and dedicated external reviewers is a tantalizing goal. 
To illustrate that this goal is at all practical, 
below we list several small experiments that we have carried out with the help of OpenAI's ChatGPT, showing that building blocks for such methodological automation are already available. Whether or not these particular tools and methods will indeed constitute elements of the solutions is yet to be determined.  

We conducted the experiments below with the general version  
of GPT 4.0 LLM, relying only on its built-in training corpus.
Additional developments may use LLMs that are pre-trained and fine-tuned with materials that are more narrowly focused on certain application domains and software development methods, and/or take advantage of plugins to enable ad-hoc access to materials and create custom functionality.

\subsubsection{From user-defined terms to domain ontologies.} 
In order to map what an application does into available domain knowledge, one of the tasks is to associate system entities with well recognized domain entities. When interpreting the text of requirements document this mapping may be more direct~\cite{tsarfaty2014semanticModelBuilding}, but what does one do when such documentation is not available or not complete? Can the source code of the application be interpreted and mapped in this manner?
We presented ChatGPT with the following prompt\footnote{We copy here our very first attempt; minor typos in the original (which ChatGPT readily overcame) were corrected here for readability of the article.} 
\textit{"In software, program code, developers try to give meaningful names to variables, functions, etc. They often use "camel case" or acronyms, or other abbreviations, like dropping vowels, etc. For example, in a financial application,
for a function that computes a final payment, an entity whose full name would be "final payment"  can be named as "FinalPayment"  or "finPayment" or "FinPmt" or "fPayment" etc.  I wonder if you can guess the full name or meaning of fields named SlsTx,  Bal, Refnd.  tntvAmt."}.The LLM readily returned \textit{Sales Tax, Balance, Refund}, and \textit{Tentative Amount}. Note that hinting about the general problem domain (finance, in this case) is not only allowed, but perhaps even desired in such analyses by humans and machines. 
Clearly, with sufficient training data from source code, and some rules and examples about how programmers abbreviate and compound words to assign names to software entities, tools can infer the ontological associations of system entities.  

\subsubsection{From computations and programmed scenarios to domain processes.}
To learn what a software application does and does not do, one needs to look at sequences of events and operations.  ChatGPT has proven useful in interpreting  computations that often call for human explanation. To our prompt \textit{``Can you guess `names' of variables,  or `their purpose' from the logic of the computation? Here is an example. Can you associate it with `a story'? 
X1= 0.2;
X5=0.06
X2=<Input>;  X3=X2-X2*X1; X4=X3*(1+X5)''}, ChatGPT readily responded that this program snippet may be computing the final price of a product by applying discounts and taxes to the base price. The same results were obtained also when such a computation was carried out in a single expression without interim variables. 

\subsubsection{Creating ontological checklists.}
While databases of curated ontologies exist, ChatGPT was able to readily retrieve compact lists of issues that must be handled in a given domain. Asking for the issues that must be handled in a rudimentary web store application it returned a long list of areas that must be handled, from creating invoices through managing stock to dealing with the fact that certain taxes depend on location. Such lists are an essential component of our intended analyses.

\subsubsection{Comparing lists.}
With components like those detailed above, one can create lists of issues and entities that are expected in the problem domain, and similar issues and entities as represented in the target system.  
Clearly, one does not need an LLM to compare well-structured lists, but it was nice to see that ChatGPT could compute set difference between two sets defined with text descriptions. In this way, one can discover problem domain issues that were not conspicuously handled in the application. The inverse can be used for enriching the ontology. 

\subsubsection{Using domain knowledge for adding function.}  
Inspired by an ecological model described in~\cite{marronEtAl2023ModelingChallenges}, we presented to ChatGPT a simple scenario-based model of the interaction of bees and flowers.
We explained to ChatGPT that this model is not part of a biological study but 
is part of a science education program that focuses on computational/algorithmic thinking skills, by describing natural phenomena as composition of compact, stand-alone rules/scenarios. The rules in the model revolved around monthly changes in patterns of average daylight hours and temperature, and the effect of these parameters on whether flowers bloom, and whether bees are active or hibernate. We asked ChatGPT to suggest additional topics that could enrich the model, while keeping with the goals: i.e., rather than adding all biological and ecological factors that could be relevant (and that ChatGPT was clearly aware of), suggest one or two other similar factors. ChatGPT suggested allowing for variations due to the changing weather that may change the temperature from the expected monthly averages. This is also an example of the potential duality between omissions and hidden assumptions: not accounting for changes in the weather may be a sheer omission or a conscious choice. This example also clarifies the importance of documenting assumptions as such: having a table for ``the temperature'' during each month does not necessarily mean that the model assumes that the temperature is fixed throughout each month---other computations may still modify the ambient temperature. Adding a sentence that states this assumption explicitly, clarifies the issue and supports a more thorough evaluation of the model.

\subsubsection{Using domain knowledge for state refinement.}  
Along the lines of Section~\ref{sec:StrucureDrivenDiscovery} we ran the following experiment. 

\textbf{Our prompt:}
`Vehicle State Refinement: In modeling complex systems, developers often establish orthogonal, alternative states. For example, in describing a vehicle's behavior or operation, one may recognize the Boolean state variables of S1 as ``moving'' vs. ``standing'', and S2 as ``engine on'' vs. ``engine off''. Then there are composite states with combinations of these state variables.
Such composite states sometimes give the illusion of full coverage of possible states,
but domain knowledge may require finer differentiation like S3 as ``vehicle is parked'' vs. ``vehicle in use'' or S4 as ``vehicle in maintenance (as in being in the shop)'' vs. ``vehicle is operational''.
Can you suggest more state refinements like these for the case of vehicles?
My goal is to show that ``broad brush'' discrete division of states is sometimes misleading, and may cause one to overlook important variables, and that domain expertise can help correct the situation'.

\textbf{ChatGPT} (pruned, with added numbering): \textit{
Absolutely, state refinements can lead to a more accurate understanding and modeling of complex systems. For a vehicle, there are indeed numerous additional states to consider that can provide a deeper understanding of the vehicle's condition and operation. Here are some examples:   (1) Fuel Level: This state could be divided into categories such as "full", "partially full", "low", or "empty"...  (2) Load: ... "empty", "partially loaded", or "fully loaded". This state has implications for fuel consumption, performance, and wear and tear on the vehicle. (3) Weather Conditions: ... "clear", "rain", "snow", "fog", etc. ... affect vehicle performance and safety ... (4) Road Conditions: ... "smooth", "bumpy", "icy", "wet", etc. ... (5) Vehicle Integrity: "Optimal", "minor damage", "major damage", "totalled"... (6) Occupancy: ..."unoccupied", "driver only", "passengers present", "at full capacity". Occupancy not only ...weight distribution,... driver behavior ... (7) Vehicle Mode: ... "electric mode", "fuel mode", or "mixed mode". Similarly, ... "manual driving", "assisted driving", or "fully autonomous" mode. (8) Maintenance Status: "Recently serviced", "due for service", "overdue for service". ... (9) Tire Condition: "New", "good", "worn", "balding", "flat". ... (10) Driver Condition:... "Alert", "fatigued", "impaired", "distracted". While not strictly a state of the vehicle, the driver's condition heavily influences the operation and safety of the vehicle.  
This list can be extended further depending on the level of granularity required and the specific use case. By identifying and considering these states, developers can model vehicles more accurately and comprehensively."}

\subsubsection{Learning from gaps in tools.}

A tool designed to help uncover hidden assumptions, can also be helpful when it misses the mark. Failures can be studied in the context of a synthetic application constructed to challenge the tool,  or by comparing the tool's output to comments by external reviewers, or as part of in a retrospective failure analysis.
In such cases, having a concise description of a hidden assumption that tool has missed, can teach us about human thinking and  development processes and guide incremental enhancements to the tool and the discovery processes.
In another experiment we asked ChatGPT for the motion formulas of a pendulum. We then asked for assumptions in this model, and ChatGPT listed thirteen assumptions, including small angle, no friction, rigid rod, point mass, and more. However, ChatGPT did not list the assumptions that the pivot point is fixed in space, and that the pendulum motion is confined to two dimensions --- a plane. A generalized enhancement that we derive from this experiment, which can be applied to other systems, is always adding questions ``Which points in this configuration are fixed in space?'', and ``In how many dimensions is the problem considered?''. With regard to the pendulum, in a separate experiment,  ChatGPT answered these questions correctly, i.e., that the pivot is fixed and the motion is in two dimensions; furthermore, in response to another question, ChatGPT admitted that these two facts really belonged in the original list of assumptions. 
A similar discussion about insights to be gained from incremental development is made in~\cite{eisenstein2023Turing} in the context of how humans develop  increasingly sophisticated Turing Tests that aim to challenge the intelligence of machines that have passed lower level tests.

\subsubsection{Human interfaces for discovery tools.}

The above experiments suggest that in designing the interfaces for tools for discovery of hidden assumptions, developers can initially suffice with mimicking the interactions with external reviewers: feed the tools with the same development and marketing documents, code artifacts, simulation outputs and videos, etc., as would be shown to reviewers, and then ask the same kind of questions. 
To make-up for the expert's knowledge and experience which may not already be built into the tool (say, as databases or AI training), additional materials from other systems, reference documents, etc. may be fed for comparison and matching.  

\section{Conclusion}

Reviews by external application specialists and technology professionals are considered of great potential value to a project. 
Here, and as part of the vision of wise computing~\cite{harel2018wiseComp}, we presented initial steps for creating automated tools that help mimic some of the critical constructive reviews that outside experts can offer. Such tools can facilitate carrying out these reviews constantly, throughout development, and from the very beginning of the project, reducing costs and improving quality. 

Future research directions include refining the definitions and types of hidden assumptions, building prototype tools, and enhancing the community-wide availability of relevant machine-accessible knowledge and expertise.

\section*{Acknowledgements}
We thank the anonymous reviewers for their valuable comments and suggestions. 
This project has received funding from the European Union’s Horizon research and innovation programme under grant agreements no 101094905 (AI4Gov).
This research was funded in part by an NSFC-ISF grant issued jointly by the  National Natural Science Foundation of China (NSFC) and the Israel Science Foundation (ISF grant 3698/21). Additional support was provided by a research grant from the Estate of Harry Levine, the Estate of Avraham Rothstein, Brenda Gruss, and Daniel Hirsch, the One8 Foundation, Rina Mayer, Maurice Levy, and the Estate of Bernice Bernath


\end{document}